\documentclass[aps,prl,twocolumn,showpacs,superscriptaddress]{revtex4-2} 

\usepackage{graphicx}
\usepackage[colorlinks,citecolor=blue,urlcolor=blue,linkcolor=blue]{hyperref}
\usepackage{siunitx}	
\usepackage{amsmath}
\usepackage{amsthm}
\usepackage{amssymb}
\usepackage{amsthm}
\usepackage{braket}
\usepackage{booktabs}
\usepackage{footmisc}
\usepackage{mathrsfs}
\usepackage{cleveref}
\usepackage{dsfont}
\usepackage{subfigure}
\usepackage{multirow}
\usepackage{enumerate}
\usepackage{lineno}
\usepackage{mathtools}
\usepackage{physics}
\usepackage{svg}

\newcommand{\bracket}[2]{\ensuremath{\langle#1 \vphantom{#2}| #2\vphantom{#1}\rangle}}

\newcommand{\com}[1]{}

\begin{document}
\title{Testing Born's rule via photoionization of helium}

\author{Peter Robert Förderer}
\affiliation{Physikalisches Institut, Albert-Ludwigs-Universit{\"a}t Freiburg, Hermann-Herder-Straße 3, 79104 Freiburg, Germany}
\affiliation{EUCOR Centre for Quantum Science and Quantum Computing, Albert-Ludwigs-Universität Freiburg, Hermann-Herder-Straße 3, 79104 Freiburg, Germany}

\author{Andreas Buchleitner}
\email{a.buchleitner@physik.uni-freiburg.de}
\affiliation{Physikalisches Institut, Albert-Ludwigs-Universit{\"a}t Freiburg, Hermann-Herder-Straße 3, 79104 Freiburg, Germany}
\affiliation{EUCOR Centre for Quantum Science and Quantum Computing, Albert-Ludwigs-Universität Freiburg, Hermann-Herder-Straße 3, 79104 Freiburg, Germany}

\author{David Busto}
\affiliation{Physikalisches Institut, Albert-Ludwigs-Universit{\"a}t Freiburg, Hermann-Herder-Straße 3, 79104 Freiburg, Germany}
\affiliation{Department of Physics, Lund University, Box 118, 22100 Lund, Sweden}

\author{Christoph Dittel}
\email{christoph.dittel@physik.uni-freiburg.de}
\affiliation{Physikalisches Institut, Albert-Ludwigs-Universit{\"a}t Freiburg, Hermann-Herder-Straße 3, 79104 Freiburg, Germany}
\affiliation{EUCOR Centre for Quantum Science and Quantum Computing, Albert-Ludwigs-Universität Freiburg, Hermann-Herder-Straße 3, 79104 Freiburg, Germany}
\affiliation{Department of Physics, Lund University, Box 118, 22100 Lund, Sweden}

\date{\today}

\begin{abstract}
It is shown how state-of-the-art attosecond photoionization experiments can test Born's rule -- a postulate of quantum mechanics -- via the so-called Sorkin test. A simulation of the Sorkin test under consideration of typical experimental noise and data acquisition efficiencies infers an achievable measurement precision in the range of the best Sorkin tests to date. The implementation of further fundamental tests of quantum mechanics is discussed.
\end{abstract}
\maketitle

Attosecond physics emerged around the beginning of this millennium, from the desire to resolve electronic dynamics in matter on short time scales \cite{Krausz2009,Calegari2016}. It encompasses the ionization of atoms \cite{Kienberger2004,Sansone2006,Bergues2012,Kaldun2016}, molecules \cite{Sansone2010,Gong2023}, and solids \cite{Cavalieri2007}, typically through the exposure to short, extreme-ultraviolet (XUV) light pulses, often synchronized with infrared (IR) probe pulses to analyze the induced photoionization dynamics \cite{Paul2001,Kotur2016,Gruson2016,Busto2018,Priebe2017,Bourassin2020,Laurell2022,Laurell2023}. The ever-increasing stability and controllability of available light sources promises the synergy of attosecond physics and quantum information science \cite{Tichy-EE-2011,Lewenstein2022}. First steps approach the exploration of entanglement in double photoionization \cite{Akoury-SD-2007,Zimmermann-LL-2008,Burkhard-LL-2009}, electron-ion entanglement \cite{Vrakking2021,Koll2022,Laurell2022,Laurell2023} or the driving of high-harmonic generation via non-classical light fields \cite{Stammer-QE-2023,Gorlach2023,Bhattacharya2023}, while implementations of quantum information processing and fundamental tests of quantum theory have not been achieved yet in these rather complex settings of light-matter interaction. 

Testing the foundations of quantum mechanics is tantamount to testing the postulates \cite{Tannoudji2019} on which quantum theory is build upon. Since no quantum mechanical prediction has so far been violated within the limits of experimental precision, there is no evidence against the validity of these postulates. Revealing the limits of quantum theory therefore requires, if at all, highly precise experiments, ideally targeted to test specific postulates in all possible physical system. One of these postulates is Born's rule \cite{Born1926}. It establishes the connection between the mathematical formalism and the experimentally recorded counting statistics by stating that event probabilities are given by the squared modulus of the associated transition amplitudes. Since Born's rule implies the restriction of quantum interference to pairs of paths \cite{Born1926},  three- or multi-path interferometers specifically allow to asses the validity of Born's rule via the so-called Sorkin test \cite{Sorkin1994}, and to rule out alternative theories \cite{Dakic2014,Zyczkowski2008}. However, none of the hitherto implemented Sorkin tests using single photons~\cite{Sinha2010,Soellner2011,Hickmann2011,Magana-Loaiza2016,Kauten2017,Vogl2021}, nuclear magnetic resonance in a spin ensemble \cite{Park2012}, a single nitrogen-vacancy center in diamond \cite{Jin2017}, atomic matter waves \cite{Barnea2018}, single dye molecules \cite{Cotter2017}, two-photon interference \cite{Pleinert2021}, and coherent light together with homodyne detection \cite{Conlon-TP-2024} (see \cite{Gstir2023} for an overview) found evidence against the validity of Born's rule. To further assess the validity and accuracy of Born's rule therefore calls for an increased precision in these experiments, as well as an expansion of the Sorkin test to physical systems that have not been considered yet for its implementation.


Here, we propose an implementation of the Sorkin test in the photoionization process of helium. We show how spectral shaping of the IR probe pulse in typical pump-probe photoionization experiments allows the realization of three-path interferometers. We perform a Monte Carlo simulation of the Sorkin test in consideration of typical noise sources in state-of-the-art experiments, and infer an achievable precision of $10^{-3}$, i.e., at the same level as most Sorkin tests to date \cite{Park2012,Kauten2017,Jin2017,Sadana2022}. The interferometer's intrinsic phase stability makes attosecond photoionization also a promising candidate for further tests of quantum mechanics, such as the so-called Peres test \cite{Peres1979}, which we briefly discuss at the end of this contribution. As a result, we establish the potential of attosecond photoionization as a platform for fundamental tests of quantum mechanics.


\textit{Sorkin test -- }Given a quantum system described by the state $\ket{\psi}$ in a Hilbert space $\mathcal{H}$, Born's rule \cite{Born1926} states that the measurement of an observable $O$ on $\mathcal{H}$, with eigenstates $\ket{\lambda_i}$ and corresponding eigenvalues $\lambda_i$, yields the outcome $\lambda_i$ with probability $P  = \abs{\bra{\lambda_i}\ket{\psi}}^2$. As a direct consequence, for superpositions $\ket{\psi}=\alpha_a \ket{\psi_a}+\alpha_b\ket{\psi_b}$ of distinct components $\ket{\psi_a}$ and $\ket{\psi_b}$, with $\alpha_j \in\mathbb{C}$, such as in a double-slit experiment, it entails the interference of the transition amplitudes $A_a=\alpha_a \bracket{\lambda_i}{\psi_a}$ and $A_b=\alpha_b \bracket{\lambda_i}{\psi_b}$: The probability $P_{ab} = \abs{A_a+A_b}^2= P_a + P_b + I_{ab}$ to observe the outcome $\lambda_i$ depends not only on the probabilities $P_j = \abs{A_j}^2$ corresponding to the individual components $j\in\{a,b\}$, but also on the interference term $I_{ab} = A_a^*A_b+ A_a A_b^*$, which is sensitive to the relative phase between $A_a$ and $A_b$. By adding a third component, $P_{abc} = \abs{A_a+A_b+A_c}^2 = P_a + P_b + P_c + I_{ab} + I_{ac} + I_{bc}$, the classical additivity of the individual probabilities $P_j$, $j\in\{a,b,c\}$, is merely accompanied by interference terms $I_{jk}=P_{jk}-P_j-P_k$ between pairs of amplitudes, and no additional interference term between more than two transition amplitudes intervenes. Consequently, following the idea of Sorkin \cite{Sorkin1994}, we can define such an additional interference term $I_{abc}$ as the deviation from the quantum mechanical prediction for $P_{abc}$ \cite{Sinha2010}, which can be expressed in terms of measurable probabilities alone, $I_{abc} = P_{abc} - P_{ab}-P_{ac}-P_{bc}+P_a+P_b+P_c$, and must vanish if Born's rule applies. In order to correct for constant background signals in experiments, one usually subtracts the detection signal $P_0$ in the absence of all components from each probability $P_S$, with $S\in\qty{a,b,c,ab,ac,bc,abc}$, i.e., $P_S'= P_S-P_0$, yielding the corrected interference terms $I_{abc}' = I_{abc}-P_0$ and $I_{jk}' = I_{jk}+P_0$ \cite{Sinha2010,Soellner2011,Park2012,Kauten2017,Lee2020,Gstir2023}, which constitute the so-called \textit{Sorkin parameter} \cite{Sinha2010}
\begin{align}\label{eq:sorkinparameter}
    \kappa &\coloneqq  \frac{I_{abc}'}{\abs{I_{ab}'} + \abs{I_{ac}'} + \abs{I_{bc}'}}.
\end{align}
It defines a quantitative measure of potential deviations \cite{Sorkin1994} from Born's rule, experimentally accessible by recording all the probabilities $P_S$, $S\in\qty{0,a,b,c,ab,ac,bc,abc}$.

\begin{figure}[t]
    \centering
    \includegraphics[]{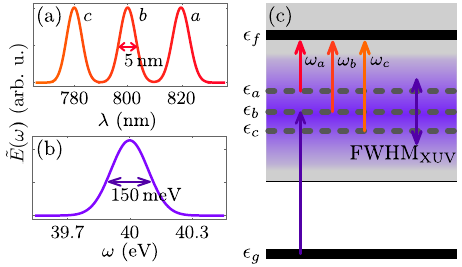}
    \caption{Spectral representation of the photoionization model. Spectrum (a) of the trichromatic IR probe pulse, with components $a$, $b$, and $c$, and (b) of the XUV pump pulse. Horizontal arrows indicate the FWHM of the laser pulses. (c) Sketch of the two-photon processes which mediate the photoionization process. The absorption of a temporally short XUV photon (violet upwards arrow) excites the electron from the bound ground state $\ket{g}$ with energy $\epsilon_g$ to the single particle ionization continuum (gray-shaded area), populating a broad distribution of scattering states (violet-shaded area with width $\mathrm{FWHM}_\mathrm{XUV}$). A subsequent absorption of a spectrally narrow single photon from either one of the three spectral components of the IR pulse (red lines) can lead to the final photoelectron energy state $\ket{\epsilon_f}$.}
    \label{fig:energydiagram}
\end{figure}

\textit{Photoionization model -- }As we show in the following, state-of-the-art experiments on the photoionization of atoms via short multi-chromatic laser pulses \cite{Gruson2016,Villeneuve2017,Barreau2019,Koll2022,Laurell2023,Luo2023} realize stable multi-path interference, which serves well for high-precision measurements of the Sorkin parameter~\eqref {eq:sorkinparameter}. To this end, consider single-electron, laser-assisted photoionization \cite{Agostini1979} into the structureless continuum of helium atoms, induced by a broadband extreme ultraviolet (XUV) pump pulse, with central frequency $\omega_\mathrm{XUV}$ and spectral width $\mathrm{FWHM}_\mathrm{XUV}$, as well as a synchronized infrared (IR) probe pulse, composed of three spectral components $j\in \qty{a,b,c}$ with central frequencies $\omega_j$, such that $|\omega_j-\omega_k|\lessapprox \mathrm{FWHM}_\mathrm{XUV}$, and equal, narrow spectral widths $\mathrm{FWHM}_j = \mathrm{FWHM}_\mathrm{IR}$ [see Figs.~\ref{fig:energydiagram}(a,b)]. Note that, in contrast to typical pump-probe spectroscopy schemes \cite{Paul2001}, here the time delays $\tau_j$ between the $j$th IR component and the XUV pump pulse are ideally kept constant during the measurement. The absorption of a photon from the XUV pulse induces a transition from the electron's bound ground state $\ket{g}$ to an intermediate continuum state with broad energy distribution (as naturally provided via femto- or attosecond laser pulses) that has significant overlap with three distinct continuum states $\ket{\epsilon_j}$ corresponding to the energies $\epsilon_j = \epsilon_f-\hbar\omega_j$, where $\ket{\epsilon_f}$ is the photoelectron's final state, interrogated by the experimental setup [see Fig.~\ref{fig:energydiagram}(c)]. Consequently, the subsequent absorption of a photon from either one of the spectrally narrow components $j\in \qty{a,b,c}$ of the IR pulse feeds the final energy state $\ket{\epsilon_f}$ by three transition amplitudes, thus realizing a three-path interferometer.

Different path-configurations $S$ can be realized through the selection of subsets of the three available IR components $j\in \qty{a,b,c}$, such that the corresponding probability to detect a photoelectron with energy $\epsilon_f$ yields $P_S$. For example, $P_{ac}$ is obtained in the presence of  the IR components $a$ and $c$, and in the absence of component $b$. In this way, one has access to all probabilities $P_S$ which enter Eq.~\eqref{eq:sorkinparameter}.

\textit{Theoretical description -- }To specify the probability $P_S = \abs{A_S}^2$, with amplitude $A_S= \sum_{j\in S}A_j$, we follow Ref.~\cite{Galan2016} and derive an explicit expression for the individual transition amplitudes $A_j$ (see \cite{Foerderer2022} for details): Due to the injected light fields' intensities and their large wavelengths as compared to the electronic initial state's scale, we can treat the laser fields classically and describe the light-matter interaction in dipole approximation, with the Hamiltonian $H_S = H_0 + V_S(t)$ \cite{Cohen-Tannoudji1998}, where $H_0$ is the field-free atomic component, and $V_S(t) = - d\,E_S(t)$ the interaction term, with $d$ the dipole operator and $E_S(t) = E_\mathrm{XUV}(t) + \sum_{j\in S} E_j(t)$ the total (phase stable) driving field, composed of the XUV pump $E_\mathrm{XUV}(t)$ and the IR spectral components $E_j(t)$, for all $j\in S$. In the weak coupling regime and in the absence of intermediate atomic resonances, the interaction term can be treated as a perturbation, with the perturbation order corresponding to the number of photons involved in the ionization process \cite{Cohen-Tannoudji1998}. For initial interaction times $t_0 \rightarrow -\infty$ and measurement times $t \rightarrow  \infty$, the amplitude $A_S$ can be expanded to second order as
\begin{align}\label{eq:Athird}
\begin{split}
    &A_S = -\frac{\mathrm{i}}{\hbar} \int_{-\infty}^{\infty}\dd{t'}\bra{\epsilon_f}\tilde{V}_S(t')\ket{g} \\
   &- \frac{1}{\hbar^2}\int_{-\infty}^{\infty}\dd{t'}\int_{-\infty}^{t'}\dd{t''}\bra{\epsilon_f}\tilde{V}_S(t')\tilde{V}_S(t'')\ket{g}+\dots,
\end{split}
\end{align}
with $\tilde{V}_S(t)$ the interaction Hamiltonian in the interaction picture with respect to $H_0$. By restricting ourselves to final energies $\epsilon_f \approx \epsilon_g+ \hbar\omega_\text{XUV}+ \hbar \omega_\text{IR}$, with $\hbar \omega_\text{IR}\gg\text{FWHM}_\text{XUV}$, in a featureless continuum [see Fig.~\ref{fig:energydiagram}(c)], any single-photon absorption process becomes off-resonant, such that the first order term in~\eqref{eq:Athird} vanishes. The first non-vanishing term in~\eqref{eq:Athird} is thus of second order. It includes all four time-ordered combinations of two-photon processes (XUV$\rightarrow$IR, IR$\rightarrow$XUV, XUV$\rightarrow$XUV, and IR$\rightarrow$IR). However, only the absorption of a XUV photon followed by the absorption of an IR photon contributes, since both transitions XUV$\rightarrow$XUV and IR$\rightarrow$IR are highly off-resonant, and the IR$\rightarrow$XUV transition is highly unlikely as compared to the reverse time ordered one due to the absence of resonant intermediate states. Any three-photon process is also off-resonant, and, thus, suppressed. The first neglected, non-suppressed term is thus of fourth order. The transition amplitudes $A_j$, $j\in \{a,b,c\}$, can therefore be written as \cite{Galan2016,Foerderer2022}
\begin{align}
\begin{split}\label{eq:transitionamplitude}
    A_j &=  \lim_{\eta\searrow  0}\frac{\mathrm{-i}}{\hbar^2} \int_{-\infty}^{\infty} \dd{\omega} \tilde{E}_j(\omega_{fg} -\omega;\tau_j) \tilde{E}_{\mathrm{XUV}}(\omega)\\
    &\hspace*{1.2cm}\times\bra{\epsilon_f}  d  \frac{1}{\omega_g + \omega - H_0/\hbar +\mathrm{i}\eta}   d  \ket{g},
\end{split}
\end{align}
with $\omega_{fg} = (\epsilon_f-\epsilon_g)/\hbar$, and $\tilde{E}_\mathrm{XUV}(\omega)$ and $\tilde{E}_j(\omega;\tau) = \mathrm{exp}(i\omega\tau_j) \tilde{E}_\mathrm{IR}(\omega)$ the electric XUV and IR field in the frequency representation, respectively. Note that the neglected non-vanishing higher-order terms can give rise to a non-vanishing Sorkin parameter without a violation of Born's rule, similar as in optical three-path experiments \cite{Sawant2014,Sinha2015,Magana-Loaiza2016,Namdar-EH-2023}.

As illustrated in Figs.~\ref{fig:energydiagram}(a,b), we model the pump and probe fields by Gaussian pulses $\tilde{E}_\mathrm{XUV}(\omega)$ and $\tilde{E}_j(\omega)$, with central frequencies $\omega_\mathrm{XUV}={\SI{40}{eV}}$ and $\omega_{a/b/c}=820/800/{\SI{780}{nm}}$, respectively, and spectral widths $\mathrm{FWHM}_\mathrm{XUV}={\SI{150}{meV}}$ and $\mathrm{FWHM}_j=\mathrm{FWHM}_\mathrm{IR}={\SI{5}{nm}}$. Since the helium target exhibits a featureless continuum between the single ionization threshold and doubly excited states at energies around \SI{60}{eV} \cite{Fano1961,Domke1996}, addressed by the here employed driving frequencies, the intermediate and final states are featureless continuum states for which we can employ the so-called on-shell approximation \cite{Galan2016}.

In general, there can be multiple ionization channels \cite{Fano1985,Busto2019}, such that photoelectrons with final energy $\epsilon_f$ can have different angular momentum states. For two-photon absorption in helium, the dipole selection rules allow the transition from the $s$ ground state over a $p$ intermediate state to either an $s$ or a $d$ final state. For the measurement of the Sorkin parameter~\eqref{eq:sorkinparameter}, these angular momentum states, however, need not be resolved, allowing for angle-integrated measurements. In order to make this explicit, consider the photoelectron's continuum basis states $\ket{\epsilon,l,m}$, with energy $\epsilon$, angular momentum quantum number $l$, and magnetic quantum number $m$. The photoelectron's state after two-photon absorption can then be written as $\ket{\psi}=\int\mathrm{d}\epsilon \sum_{j\in S\setminus \{0\}} [\alpha_{j,0,0}(\epsilon) \ket{\epsilon,0,0} + \alpha_{j,2,0}(\epsilon) \ket{\epsilon,2,0}]$, with coefficients $\alpha_{j,l,m}(\epsilon) \in \mathbb{C}$. By the orthogonality of the angular momentum states, i.e., $\bracket{l,m}{l',m'}=\delta_{l,l'}\delta_{m,m'}$, the probability to find the photoelectron in the final energy $\epsilon_f$ decomposes as $P_S=P_{S,0,0}+P_{S,2,0}$, with $P_{S,l,m}=|\sum_{j\in S\setminus \{0\}} \alpha_{j,l,m}(\epsilon_f)|^2$. Together with the linearity of the interference term $I_{abc}$ with respect to $P_S$ [see above Eq.~\eqref{eq:sorkinparameter}], we see that the photoelectron's angular momentum states have no effect on the vanishing of the Sorkin parameter~\eqref{eq:sorkinparameter}. 

\begin{figure*}[t]
  \includegraphics[width=\textwidth]{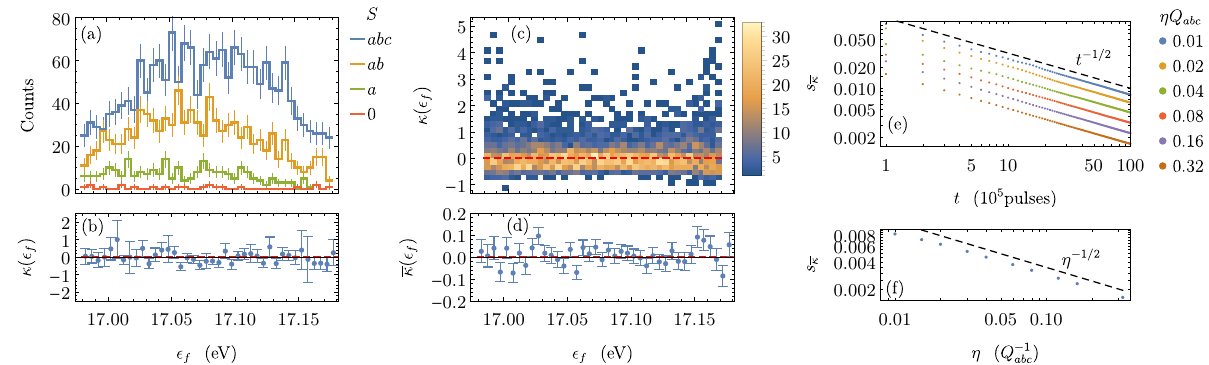}
  \caption{Monte Carlo simulation of the Sorkin experiment with photoelectrons. (a) Examples of simulated detector click statistics after $10^5$ laser pulses with typical experimental noise (see main text for details) for the path-configurations $S=abc,ab,a$, and~$0$. Error bars indicate the Poisson error for the detector click statistics. Similar results are obtained for the other path-configurations (not shown). (b) Sorkin parameter $\kappa(\epsilon_f)$ as a function of the final energy $\epsilon_f$ obtained from the data partially presented in (a). (c) Spectrally resolved abundance of $100$ Sorkin parameters $\kappa(\epsilon_f)$ per energy bin. (d) Weighted arithmetic mean $\overline{\kappa}(\epsilon_f)$ over the distribution from (c). The weighted arithmetic mean across all energies in (d) yields $\overline{\kappa}=0.0063(63)$. The red dashed line in (b-d) indicates $\kappa=0$. (e,f) Standard error $s_{\overline{\kappa}}$ vs. measurement time $t$ (e) and data acquisition efficiency $\eta$, after the recording of $100\times 10^5$ pulses (f), following a power law with exponent $-1/2$ (black dashed lines). The data point for $\eta=0.02\,Q_{abc}^{-1}$ in (f), and for $\eta=0.02\,Q_{abc}^{-1}$ and $t=100\times 10^5\,\mathrm{pulses}$ in (e), corresponds to the data presented in (c,d).}
  \label{fig:sorkinparameter}
\end{figure*}

\textit{Simulation -- }
With the help of Eq.~\eqref{eq:transitionamplitude} (see \cite{Foerderer2022} for details), we numerically simulate the Sorkin test and estimate the achievable precision of the Sorkin parameter~\eqref{eq:sorkinparameter} under typical experimental conditions for $n=40$ final energies $\epsilon_f$ in the energie window  $[\epsilon_c-\Delta\epsilon/2 , \epsilon_c+\Delta\epsilon/2]$ of width $\Delta\epsilon=\SI{0.2}{eV}$ around $\epsilon_c=\epsilon_g+\hbar(\omega_\mathrm{XUV}+\omega_b)\approx \SI{17.08}{eV}$, specified by  $\epsilon_f=\epsilon_c-\Delta\epsilon/2+\Delta\epsilon(f+1/2)/n$, with $f=0,\dots, n-1$. To this end, we conservatively estimate the most dominant sources of noise in typical state-of-the-art experiments. For the field amplitudes $\tilde{E}_\mathrm{XUV}$ and $\tilde{E}_j$ of the XUV and IR light pulses [spectra shown in Figs.~\ref{fig:energydiagram}(a,b)], we consider uncorrelated, normal distributed fluctuations with standard deviations $\Delta \tilde{E} = \tilde{E}/10$. Fluctuations in the time delays $\tau_j= \tau +\tau_j'$ between the XUV pulse and the IR components are considered to arise due to normal distributed fluctuations in the common time delay $\tau$ between XUV and IR pulse with standard deviation $\Delta \tau = \SI{50}{as}$ (typical for actively stabilized interferometric pump-probe setups \cite{Luo2023}), as well as due to individual, normal distributed fluctuations in the time delays $\tau_j'$ of the $j$th IR pulse component with respect to the common IR beam (due to noise after the selection of the IR frequency components) with standard deviation $\Delta \tau_j' = \SI{200}{as}$ \cite{Luo2023}.

To account for experimental data acquisition, we perform a Monte Carlo simulation: First, we set the experiment's efficiency $\eta$ such that for the path-configuration $S=abc$ about $0.02$ electrons per laser pulse are expected in the considered energy window \cite{Luo2023}. To this end, we define $\eta = 0.02/Q_{abc}$, with $Q_{abc}=\sum_f P^{\mathrm{ideal}}_{abc}(\epsilon_f)$, where $P^{\mathrm{ideal}}_{abc}(\epsilon_f)$ is the transition probability calculated via the transition amplitudes~\eqref{eq:transitionamplitude} in the absence of noise for the final energy $\epsilon_f$, and $S=abc$. Background-signal counts are considered to appear independently of the final energy with an expected rate of $n P_0=2\times 10^{-4}$ per laser pulse in the considered energy window. For each path-configuration $S$, we then generate a random set of laser parameters $\{\tilde{E}_\mathrm{XUV},\tilde{E}_j,\tau_j\}$, calculate the corresponding detection probabilities $\widetilde{P}_S(\epsilon_f)=P_0 +\eta P_S(\epsilon_f)$ for all final energies $\epsilon_f$, and generate a single random sample from the distribution $\{\widetilde{P}_S(\epsilon_1),\dots, \widetilde{P}_S(\epsilon_n), 1-\sum_f \widetilde{P}_S(\epsilon_f) \}$, with the last element corresponding to no click in the considered energy window. By repeating this procedure $10^5$ times (i.e., for $10^5$ laser pulses, taking approximately \SI{33}{s} with a \SI{3}{kHz} laser \cite{Luo2023}), we obtain a histogram for each path-configuration $S$ that simulates the detector click statistics in the considered energy window [Fig.~\ref{fig:sorkinparameter}(a)]. From these, we extract the Sorkin parameter $\kappa(\epsilon_f)$ [see Eq.~\eqref{eq:sorkinparameter}] as a function of the final energy $\epsilon_f$ and calculate its precision under consideration of a Poisson error for the detector click statistics [Fig.~\ref{fig:sorkinparameter}(b)]. Repeating this procedure (i.e., the extraction of the spectrally resolved Sorkin parameter) $100$ times results in a distribution of the Sorkin parameter as a function of the final energy [Fig.~\ref{fig:sorkinparameter}(c)]. Note that for a typical laser repetition frequency of \SI{3}{kHz} \cite{Luo2023}, the acquisition of $100$ spectrally resolved values of $\kappa(\epsilon_f)$ (each resulting from the accumulation of $8\times 10^5$ laser pulses) takes about \SI{7.4}{h}, which is compatible with typical experimentally achievable temporal stabilities \cite{Luo2023}.

Figure~\ref{fig:sorkinparameter}(b) shows that the expected precision of a single spectrally resolved Sorkin parameter $\kappa(\epsilon_f)$ is typically of the order $10^{-1}$. When taking the weighted arithmetic mean $\overline{\kappa}(\epsilon_f)$
\footnote{Note that, due to correlations between the numerator and denominator in the definition of the Sorkin parameter from Eq.~\eqref{eq:sorkinparameter}, a bias towards a non-vanishing number can occur when averaging over multiple values of the Sorkin parameter \cite{Kauten2017}. In order to avoid this bias, we calculate the weighted arithmetic mean $\overline{\kappa}$ by taking the weighted arithmetic means of the numerator and of the denominator in~\eqref{eq:sorkinparameter} separately.}, 
over all $100$ generated values of $\kappa(\epsilon_f)$, the standard error $s_{\overline{\kappa}(\epsilon_f)}$, i.e., the precision of the Sorkin parameter, improves by an order of magnitude [Fig.~\ref{fig:sorkinparameter}(d)], since $s_{\overline{\kappa}(\epsilon_f)}$ scales with the inverse square root of the number of values of $\kappa(\epsilon_f)$. By additionally taking the weighted arithmetic mean over the final energies $\epsilon_f$ \cite{Note1}, we ultimately obtain the Sorkin parameter $\overline{\kappa}=0.0063(63)$ with an expected precision $s_{\overline{\kappa}}$ of the order $10^{-3}$.

Finally, we assess the precision of the expected Sorkin parameter as a function of the measurement time $t$ and of the experimental data acquisition efficiency $\eta$. As evident from the linear behavior on the double-logarithmic scale of Figs.~\ref{fig:sorkinparameter}(e,f), the standard error $s_{\overline{\kappa}}$ follows a power law both with respect to the measurement time (provided in units of measured laser pulses), $s_{\overline{\kappa}}\propto t^{-1/2}$, and data acquisition efficiency, $s_{\overline{\kappa}}\propto \eta^{-1/2}$. In the former case, the exponent $-1/2$ is due to the scaling of the standard error under repeated measurements, and in the latter case due to the Poisson error of the detector click statistics, which scales with the inverse square root of the number of detected events.

\textit{Discussion -- } The proposed implementation of three-path interference in the energy domain benefits from the intrinsic stability of the paths' relative phases, where phase noise is due to noise in the well-controllable XUV and IR laser pulses. At the same time, by adjusting the IR spectral components $\omega_j$ and their time delays $\tau_j$, it is possible to change the relative phases between the paths, and, hence, to systematically examine the Sorkin parameter as a function of these relative phases.

The inherent phase stability makes the proposed photoionization-based three-path interferometer also a promising candidate for the Peres test \cite{Peres1979}, which has only rarely been implemented \cite{Sadana2022,Conlon-TP-2024} due to its sensitivity to instabilities in the relative phases between the paths \cite{Gstir2021}. It examines the exclusion of quaternionic quantum mechanics \cite{Adler-QQ-1995} (a reformulation of quantum mechanics via quaternionic instead of complex numbers) through the measurement of the so-called Peres parameter $F$, with $F<1$ indicating the admissibility of quaternionic quantum mechanics, while complex quantum mechanics is admissible if $F=1$, and $F>1$ implies the violation of the superposition principle \cite{Peres1979}. Similar to the Sorkin parameter $\kappa$, the Peres parameter $F$ is a function of the probabilities $P_S$ corresponding to different path-configurations $S$ 
\footnote{In particular, $F=\alpha^2+\beta^2+\gamma^2-2\alpha\beta\gamma$, where $\alpha=(P_{bc}-P_b-P_c)/2\sqrt{P_bP_c}$,  $\beta=(P_{ac}-P_a-P_c)/2\sqrt{P_aP_c}$, and $\gamma=(P_{ab}-P_a-P_b)/2\sqrt{P_aP_b}$  \cite{Peres1979}.},
and can thus be computed from the same data set from which we obtained Figs.~\ref{fig:sorkinparameter}(c,d), resulting in the mean Peres parameter $\overline{F}=0.981(5)$
 \footnote{For each path-configuration, we sum all counts sampled from $100\times 10^5$ laser pulses with data acquisition efficiency $\eta=0.02/Q_{abc}$, and subtract the simulated dark counts. We then calculate the spectrally resolved Peres parameter $F(\epsilon_f)$ (see \cite{Note2}), and finally calculate the weighted arithmetic mean $\overline{F}$ over all final energies $\epsilon_f$.}.
Since all our simulations were performed according to the rules of standard (complex) quantum mechanics,
the slight deviation from unity suggests that reduced experimental noise (as compared to the noise levels here 
assumed) will be required to establish the Peres test as an experimental benchmark with discriminative power.

\begin{acknowledgments}
The authors are grateful to Anne L'Huillier for fruitful discussions. D.B and C.D. acknowledge financial support from the Knut and Alice Wallenberg Foundation through the Wallenberg Centre for Quantum Technology (WACQT). C.D. acknowledges financial support from the European Research Council (Advanced grant QPAP), and support by the Georg H. Endress foundation. D.B. acknowledges support from the Swedish Research Council grant 2020-06384.
\end{acknowledgments}


%

\end{document}